\input harvmac
\input amssym.tex

\lref\BerkovitsBT{
N.~Berkovits,
{\sl A new approach to superstring field theory,}
Fortsch.\ Phys.\  {\bf 48}, 31 (2000)
[arXiv:hep-th/9912121].
}
\lref\OdaBG{
I.~Oda and M.~Tonin,
{\sl On the b-antighost in the pure spinor quantization of superstrings,}
arXiv:hep-th/0409052.
}
\lref\howe{
P.~S.~Howe,
{\sl Pure spinors, function superspaces and supergravity theories in
ten-dimensions and eleven-dimensions,}
Phys.\ Lett.\ B {\bf 273}, 90 (1991).
}
\lref\BershadskySR{
M.~Bershadsky and V.~Sadov,
{\sl Theory of Kahler gravity,}
Int.\ J.\ Mod.\ Phys.\ A {\bf 11}, 4689 (1996)
[arXiv:hep-th/9410011].
}
\lref\BerkovitsTN{
N.~Berkovits,
{\sl Manifest electromagnetic duality in closed superstring field theory,}
Phys.\ Lett.\ B {\bf 388}, 743 (1996)
[arXiv:hep-th/9607070].
} 
\lref\BerkovitsWJ{
N.~Berkovits and C.~M.~Hull,
{\sl Manifestly covariant actions for D = 4 self-dual Yang-Mills and D = 10
super-Yang-Mills,}
JHEP {\bf 9802}, 012 (1998)
[arXiv:hep-th/9712007].
}
\lref\GrassiIH{
P.~A.~Grassi and L.~Tamassia,
{\sl Vertex operators for closed superstrings,}
JHEP {\bf 0407}, 071 (2004)
[arXiv:hep-th/0405072].
}
\lref\MarinoUF{
M.~Marino,
{\sl Chern-Simons theory and topological strings,}
arXiv:hep-th/0406005.
}
 \lref\BerkovitsZK{
N.~Berkovits,
{\sl ICTP lectures on covariant quantization of the superstring,}
arXiv:hep-th/0209059.
}
\lref\BI{
N.~Berkovits and V.~Pershin,
{\it Supersymmetric Born-Infeld from the pure spinor formalism of the open superstring,}
JHEP {\bf 0301}, 023 (2003)
[arXiv:hep-th/0205154].
}
\lref\ZwiebachIE{
B.~Zwiebach,
{\sl Closed string field theory: Quantum action and the B-V master equation,}
Nucl.\ Phys.\ B {\bf 390}, 33 (1993)
[arXiv:hep-th/9206084].
}
\lref\VerlindeKU{
E.~Verlinde and H.~Verlinde,
{\it A Solution Of Two-Dimensional Topological Quantum Gravity,}
Nucl.\ Phys.\ B {\bf 348}, 457 (1991).
}
\lref\Belop{
 A.~Belopolsky,
{\sl De Rham cohomology of the supermanifolds and superstring BRST  cohomology,}
Phys.\ Lett.\ B {\bf 403}, 47 (1997)
[arXiv:hep-th/9609220];\hfill\break
A.~Belopolsky, {\sl New geometrical approach to superstrings,}
arXiv:hep-th/9703183;\hfill\break
A.~Belopolsky, {\sl Picture changing operators in supergeometry and superstring theory,}
arXiv:hep-th/9706033.
}
\lref\AG{
L.~Anguelova and P.~A.~Grassi,
{\sl Super D-branes from BRST symmetry,}
JHEP {\bf 0311} (2003) 010
[arXiv:hep-th/0307260].
}
\lref\dvvtop{
 R.~Dijkgraaf, H.~Verlinde and E.~Verlinde,
{\sl Notes On Topological String Theory And 2-D Quantum Gravity,}
PUPT-1217
{\it Based on lectures given at Spring School on Strings and Quantum Gravity, Trieste, Italy, Apr 24 - May 2, 1990 and at Cargese Workshop on
Random Surfaces, Quantum Gravity and Strings, Cargese, France, May 28 - Jun 1, 1990}
}
\lref\DistlerAX{
J.~Distler,
{\it 2-D Quantum Gravity, Topological Field Theory And The Multicritical Matrix Models,}
Nucl.\ Phys.\ B {\bf 342}, 523 (1990).
}
 \lref\rfIengo{
R.~Iengo,
{\sl Computing the R**4 term at two super-string loops,}
JHEP {\bf 0202}, 035 (2002)
[arXiv:hep-th/0202058];\hfill\break
R.~Iengo and C.~J.~Zhu,
{\sl Explicit modular invariant two-loop superstring amplitude relevant for R**4,}
JHEP {\bf 9906}, 011 (1999)
[arXiv:hep-th/9905050].
}
\lref\GreenTV{
M.~B.~Green and M.~Gutperle,
{\sl Effects of D-instantons},
Nucl.\ Phys.\ B {\bf 498}, 195 (1997)
[arXiv:hep-th/9701093].
}
\lref\BerkovitsEX{
N.~Berkovits and C.~Vafa,
{\sl Type IIB R**4 H**(4g-4) conjectures,}
Nucl.\ Phys.\ B {\bf 533}, 181 (1998)
[arXiv:hep-th/9803145].
}
 \lref\BerkovitsBF{
N.~Berkovits,
{\sl A new description of the superstring,}
arXiv:hep-th/9604123.
}
\lref\NekrasovJS{
N.~Nekrasov, H.~Ooguri and C.~Vafa,
{\it S-duality and topological strings,}
JHEP {\bf 0410}, 009 (2004)
[arXiv:hep-th/0403167].
}
\lref\GreenAS{
M.~B.~Green, M.~Gutperle and P.~Vanhove,
{\sl One loop in eleven dimensions,}
Phys.\ Lett.\ B {\bf 409}, 177 (1997)
[arXiv:hep-th/9706175].
}
\lref\HitchinJD{
N.~Hitchin,
{\it The geometry of three-forms in six and seven dimensions,}
 J. Differential Geom. {\bf 55} (2000), no. 3, 547--576
 arXiv:math.dg/0010054.
}
\lref\WittenZZ{
E.~Witten,
{\sl Mirror manifolds and topological field theory,}
arXiv:hep-th/9112056.
}
 \lref\BookTop{K. Hori et al.
{\sl Mirror Symmetry}, Clay Mathematics Monographs, V. 1, 2003}
 \lref\HitchinRW{
N.~Hitchin,
{\it Stable forms and special metrics,}
Global differential geometry: the mathematical legacy of  Alfred Gray (Bilbao, 2000),
Contemp. Math. {\bf 288}, (2001), 70--89. 
arXiv:math.dg/0107101.
}
\lref\AisakaSD{
Y.~Aisaka and Y.~Kazama,
{\it A new first class algebra, homological perturbation and extension of pure
spinor formalism for superstring,}
JHEP {\bf 0302}, 017 (2003)
[arXiv:hep-th/0212316].
}
\lref\ShatashviliZW{
S.~L.~Shatashvili and C.~Vafa,
{\it Superstrings and manifold of exceptional holonomy,}
arXiv:hep-th/9407025.
}
\lref\IqbalDS{
A.~Iqbal, N.~Nekrasov, A.~Okounkov and C.~Vafa,
{\it Quantum foam and topological strings,}
arXiv:hep-th/0312022.
}
\lref\WittenFB{
E.~Witten,
{\sl Chern-Simons gauge theory as a string theory,}
Prog.\ Math.\  {\bf 133}, 637 (1995)
[arXiv:hep-th/9207094].
} 
\lref\SamuelFE{
S.~Samuel,
{\sl Solving The Open Bosonic String In Perturbation Theory,}
Nucl.\ Phys.\ B {\bf 341}, 513 (1990).
}
\lref\GrassiTZ{
P.~A.~Grassi, G.~Policastro and P.~van Nieuwenhuizen,
{\sl The massless spectrum of covariant superstrings,}
JHEP {\bf 0211}, 001 (2002)
[arXiv:hep-th/0202123].
}
\lref\BershadskyCX{
M.~Bershadsky, S.~Cecotti, H.~Ooguri and C.~Vafa,
{\sl Kodaira-Spencer theory of gravity and exact results for quantum string
amplitudes,}
Commun.\ Math.\ Phys.\  {\bf 165}, 311 (1994)
[arXiv:hep-th/9309140].
} 
\lref\AGV{
L.~Anguelova, P.~A.~Grassi and P.~Vanhove,
{\sl Covariant one-loop amplitudes in D = 11,}
arXiv:hep-th/0408171.
}
\lref\AGVtwo{L. Anguelova, P.A. Grassi and P. Vanhove, {\sl work in progress}.} 
\lref\BerkovitsRB{
N.~Berkovits,
{\sl Covariant quantization of the superparticle using pure spinors,}
JHEP {\bf 0109}, 016 (2001)
[arXiv:hep-th/0105050].
} 
\lref\rfChesterman{M.~Chesterman,
{\sl Ghost constraints and the covariant quantization of the superparticle  in ten dimensions,}
JHEP {\bf 0402}, 011 (2004)
[arXiv:hep-th/0212261].\hfill\break
M.~Chesterman,
{\sl On the cohomology and inner products of the Berkovits superparticle and
superstring,}
arXiv:hep-th/0404021.
}
\lref\rfHenneauxBook{M.~Henneaux and C.~Teitelboim,
{\sl Quantization of gauge systems,}
Princeton, USA: Univ. Pr. (1992)
}
\lref\multi{
N.~Berkovits,
{\sl Multiloop amplitudes and vanishing theorems using the pure spinor formalism
for the superstring,}
arXiv:hep-th/0406055.
}
\lref\BerkovitsUC{
N.~Berkovits,
{\sl Covariant quantization of the supermembrane,}
JHEP {\bf 0209}, 051 (2002)
[arXiv:hep-th/0201151].
}
\lref\BerkovitsFE{
N.~Berkovits,
{\sl Super-Poincare covariant quantization of the superstring,}
JHEP {\bf 0004}, 018 (2000)
[arXiv:hep-th/0001035];
}
\lref\nnekra{Talk by N.Nekrasov at Strings 2004, 
{ http://strings04.lpthe.jussieu.fr//program.php}
}
\lref\GreenKV{
M.~B.~Green, H.~h.~Kwon and P.~Vanhove,
{\sl Two loops in eleven dimensions,}
Phys.\ Rev.\ D {\bf 61}, 104010 (2000)
[arXiv:hep-th/9910055].
}
\lref\GrassiKQ{
P.~A.~Grassi, G.~Policastro and P.~van Nieuwenhuizen,
{\sl The quantum superstring as a WZNW model,}
Nucl.\ Phys.\ B {\bf 676}, 43 (2004)
[arXiv:hep-th/0307056].
}
\lref\GrassiUG{
P.~A.~Grassi, G.~Policastro, M.~Porrati and P.~Van Nieuwenhuizen,
{\sl Covariant quantization of superstrings without pure spinor constraints,}
JHEP {\bf 0210}, 054 (2002)
[arXiv:hep-th/0112162].
}
\lref\GrassiXF{
P.~A.~Grassi, G.~Policastro and P.~van Nieuwenhuizen,
{\sl The covariant quantum superstring and superparticle from their classical actions,}
Phys.\ Lett.\ B {\bf 553}, 96 (2003)
[arXiv:hep-th/0209026].
}
\lref\GrassiSR{
P.~A.~Grassi, G.~Policastro and P.~van Nieuwenhuizen,
{\sl Yang-Mills theory as an illustration of the covariant quantization of
superstrings,}
arXiv:hep-th/0211095.
}
\lref\DijkgraafTE{
R.~Dijkgraaf, S.~Gukov, A.~Neitzke and C.~Vafa,
{\sl Topological M-theory as Unification of Form Theories of Gravity,}
arXiv:hep-th/0411073.
}
\lref\GerasimovYX{
A.~A.~Gerasimov and S.~L.~Shatashvili,
{\sl Towards integrability of topological strings. I: Three-forms on
Calabi-Yau manifolds,}
arXiv:hep-th/0409238.
}
\def\newdate{22/12/2004}

\def\a{\alpha}
\def\b{\beta}
\def\g{\gamma}
\def\l{\lambda}

\def\t{\theta}

\def\G{\Gamma}

\def\p{\partial}

\Title{\vbox{
\hbox{CERN-PH-TH/2004-204}
\hbox{SPHT-T04/41}
\hbox{YITP-SB-04-62}
\hbox{hep-th/0411167}
}}
{\vbox{\centerline{
Topological M  Theory from Pure Spinor Formalism
}
}}
\smallskip
\medskip\centerline{
{\bf Pietro Antonio Grassi}$^{a,b,c,1}$ 
{and} {\bf Pierre Vanhove}$^{d,2}$}
\bigskip
\centerline{\it $^{a}$ CERN, Theory Division, 1121 Geneva 23, Switzerland,}  
\centerline{\it $^{b}$ YITP, SUNY, Stony Brook, NY 11794-3840, USA,} 
\centerline{\it $^{c}$ DISTA, Universit\`a del Piemonte Orientale,}
\centerline{\it ~~~~ Piazza Amborsoli, 1 15100 Alessandria, Italy}
\centerline{\it $^{d}$      
CEA/DSM/SPhT, URA au CNRS, CEA/Saclay,} 
\centerline{\it F-91191 Gif-sur-Yvette, France}
\centerline{${}^{1}$ {\tt pietro.grassi@cern.ch}, ${}^{2}$ {\tt pierre.vanhove@cea.fr}}
\bigskip
\bigskip

\noindent
We construct multiloop superparticle  amplitudes in 11d using the pure spinor formalism.
We explain how this construction emerges in the superparticle limit of the multiloop pure spinor superstring amplitudes prescription.
We then argue that this construction points to  
some evidence for the existence of a topological M 
theory based on a relation between the ghost number of the 
full-fledged supersymmetric critical models and the dimension 
of the spacetime for topological models. In particular, we show that 
the extensions at higher orders of the  previous results for the 
tree and one-loop level expansion for the superparticle in 11 dimensions 
is related to a topological model in 7 dimensions.

\medskip
\Date{\newdate}
\listtoc
\writetoc


\newsec{Introduction}

We learned from~\refs{\GreenAS,\GreenKV}
that stringy and membrane corrections to 11d supergravity can be captured 
by the superparticle limit of superstring or supermembrane. This was confirmed 
by the recent work \refs{\AGV} where the covariant quantized version of superparticle with the  
method of pure spinors \refs{\BerkovitsUC} was employed. 
However, that work was limited to tree and one-loop 
analysis and the measure for  such 11d superparticle amplitudes was discussed. 
Importantly it was remarked in \AGV\ that the {\it full}\foot{We mean that not only the cubic 
and quartic couplings necessary for the linear supersymmetry are correctly described by 
3-point and 4-point amplitudes in this theory, but as well the non linear terms needed for 
covariant answer. For instance from the 3 gravitons scattering one can complete the linearized 
equation of motion derived in \refs{\BerkovitsUC}. Details will be given elsewhere
\AGVtwo.}  two derivatives effective action for the 11d supergravity 
can be obtained from the Chern-Simons action 
\eqn\eCSEleven{
S_{M-th} = \int \langle U^{(3)} Q U^{(3)}\rangle + \langle U^{(3)} [U^{(1)}, U^{(3)}] \rangle + \cdots
 }
where $\langle \cdots \rangle$ is a bracket defined with the tree-level measure  from the highest scalar element in 
the (restricted) zero momentum cohomology\foot{The abbreviation $p.s.$ 
stands for ``pure spinors'' and reminds that the cohomology is 
computed in the restricted functional space (see below for the explicit form of the constraint).  
} group $H^{(7)}(Q| p.s.)$ for 
Berkovits' Pure Spinor formalism \refs{\BerkovitsUC,\AGV} with the normalization
\eqn\eTreeEleven{
\langle \lambda^{7}\theta^{9}\rangle = 1 \,,
}
where $\l^{A}$ are the 11d pure spinors (see below for their definition) and $\t^{A}$ are 
11d Majorana spinors of $Spin(1,10)$.  This formula states that the ghost number of the scalar measure for 
$[{\cal D}\l]$ is $+16$ \refs{\AGV,\multi}. More generally the total ghost number for the measure of integration
${\cal D}\l$ is the sum of the ghost number and the fermion number of the vacuum defined by $\langle \l^{p}\t^n\rangle=1$.

The pure spinor approach is based on a BRST operator $Q=\lambda^{A} d_{A}$ such that $Q^{2}= P_{M}\, (\lambda \Gamma^M \lambda)$ and $\lambda^{A}$
a commuting spinor. $d_{A}$ is the fermionic constraint for the 11d Brink-Schwarz superparticle 
\eqn\action{
S = \int d\tau \Big( P_{M} \Pi^{M} - {e\over 2} P_{M} P^{M}\Big)\,,
} 
where $\Pi^{M} = \dot x^{M} + {i\over 2} \t \G^{M} \dot \t$ 
is the supersymmetric line element and $P^{M}$ is the conjugate momentum 
to the bosonic coordinate $x^{M}$. Together with $\t^{A}$, they form the 
coordinates of 11d superspace. The Dirac matrices $\G^{M}_{AB}$ are symmetric and 
real and satisfy the Fierz identities $(\G_{M})_{(AB} \, (\G^{MN})_{_{CD)}} =0$.  

The BRST operator $Q$ squares to zero modulo the reducible constraints
\eqn\ePs{
\lambda^{A} \, (\Gamma^{M})_{AB} \lambda^B= 0, 
\quad {\rm with} \cases{m=0,\cdots 9& for d=10\cr
m=0,\cdots,9,  11 & for d=11}
}
The quantized theory has the gauged fixed action \BerkovitsRB
\eqn\actionB{
S = \int d\tau \Big( P_{M} \dot x^{M} - {1\over 2} P_{M} P^{M} + p_{A} \dot \t^{A}  + 
w_{A} \dot \l^{A} \Big)\,,
}
where $p_{A} = d_{A} - {i\over 2} P_{M} (\G^{M}\t)_{A}$ and $w_{A}$ is 
the conjugate momentum of $\l^{A}$. The action is invariant under the 
gauge transformation $\delta w_{A} = \zeta_{M} (\G^{M})\l_{A}$ generated 
by the pure spinor constrains \ePs.  

The physical states are identified by the BRST cohomology and for 
our purposes we are interested in two types of cohomologies: 
for $P_{M} \neq 0$ (where the only non-vanishing cohomologies 
$H^{(3)}(Q|p.s.) \simeq H^{(4)}(Q|p.s.) \neq 0$ and $H^{(n)}(Q|p.s.) =0$ 
for $n\neq 3,4$) and the zero momentum cohomology for $P_{M}=0$ (where $H^{(n)}(Q|p.s.)\neq 0$ for $0 \leq n \leq 7$). 
In the latter case, $Q^{2}=0$, since $P_{M}=0$, but we still define the cohomology 
as the constrained cohomology. Actually  the (restricted) zero momentum cohomology group $H^{(*)}(Q| p.s.)$
contains all the fields of 11d supergravity,  
the ghosts, the ghost-for-ghosts and the antifields for their symmetries 
 \refs{\BerkovitsUC,\BerkovitsFE,\rfChesterman}.

The pure spinor constraints \ePs\ are reducible and the cohomology 
is best studied by introducing new ghosts at each level of reducibility and
redefining the BRST operators $Q$. 
This amounts to relaxing the constraints and replaced them 
by new terms in the BRST charge.  This approach has been pursued 
and developed in \refs{\GrassiUG,\GrassiTZ,\GrassiXF,\GrassiKQ}; there a suitable 
treatment of the ghost-for-ghost  system is obtained by introducing  a new quantum 
number (the grading) and requiring that physical states are in a 
restricted functional space. Furthermore, in \rfChesterman, it is shown that a 
straightforward application of the Homological Perturbation Theory 
techniques (see for example refs.~\rfHenneauxBook, 
and \refs{\GrassiSR, \AisakaSD} 
for the application to string theory) leads to an infinite set of ghost-for-ghosts and the cohomology $H^{(*)}(Q|p.s.)$ is 
obtained as a relative cohomology $H^{(*)}(Q,H^{(*)}(Q'))$ of a second BRST 
charge $Q'$. This charge implements the constraints at the quantum level.

Analogously to 11d superparticle, one can study 10d SYM theory or 
N=2 10d supergravities as the zero slope approximation of 
the open/closed superstrings.    
Denoting by $Q_{o}, Q_{L/R}$ and $Q$ the BRST operators for the 
open, the closed superstrings  and 
for the supermembrane or their respective superparticle limit, one finds that 
the  zero momentum cohomology  for the case of open/closed superstring 
\BerkovitsFE\  
and the supermembrane  \BerkovitsUC\ 
reveals that the highest element is contained 
in the groups $H^{(3)}(Q_{o}|p.s.)$, 
$H^{(3)}_{L}(Q_{L}|p.s.) \otimes H^{(3)}(Q_{R}|p.s.)$ and 
$H^{(7)}(Q|p.s.)$, respectively. 

A multiloop prescription for the closed string was constructed in \multi, whose particle limit
leads to a correct prescription 
for higher loop computations in quantum field theory using 
world-line formalism.
Here we construct the measure of integration for all higher-loop amplitudes for the 11d superparticle.

One of the major difficulties for a prescription that works 
for higher loop expansion is that the model is supposed 
to describe a theory of supergravity. Indeed, by a simple dimensional 
reduction it reduces to 10d type IIA superparticle. The latter is 
a particle limit of superstring quantized with the pure spinors constraints. 
The construction of multiloop formalism can be done by 
analogy with superstring prescription of \multi. The measure that 
we are looking for has to fulfill the following requirements; {\it 1)} saturation of 
ghost zero modes for $\l^{\a}$ (those zero modes are bosonic and therefore 
Dirac delta functions are needed to avoid divergences; those delta functions 
are indeed present into the picture changing operators); {\it 2)} saturation of 
bosonic ghost $w_{A}$, (again those zero modes are absorbed by the 
delta functions of picture changing operators); {\it 3)} the zero modes of 
the fermionic $d_{A}$'s which has to saturate the Berezin integration and finally, 
{\it 4)} the number of zero modes of $\t^{A}$ that have to select the correct term 
of the effective action matching the various non-renormalization theorems for higher-derivative terms. 

These requirements imply that 
the number of insertions of ``anti-ghost'' $b$-field  has to be equal to $6(g-1)$ for the 10d superparticle, and 
 we will find that 
the correct number of insertions of $b$-field has to be $7(g-1)$ which 
seems to suggest a corresponding number 
of moduli. In addition, the number of picture changing operators are 
accordingly obtained. 
This counting of moduli and the number of insertions match with those 
of a topological model in 7 dimensions. This existence of this 
model has been conjectured earlier (see the talk by N. Nekrasov 
 \nnekra) and 
 a proposal was recently formulated by Gerasimov and Shatashvili \GerasimovYX\ (see as well \refs{\DijkgraafTE}). 

We  argued that all the topological theories can be derived 
from the pure spinor approach to open/closed Superstring Field Theory  \refs{\BerkovitsBT} and 11d action \refs{\AGV} by a consistent reduction 
(see the Table).  

\begingroup
$$\vbox{\halign{ \hfil#\hfil\quad&\hfil#\hfil\quad&\hfil#\hfil\quad&\hfil#\hfil\cr
\noalign{\hrule\vskip .5mm \hrule}\cr
highest state& dimension &ghost anomaly & top model\cr
\noalign{\smallskip \hrule}\cr
$\lambda^{3}\theta^5$ &$d=3$& $-8$ & open string\cr
$\lambda^{6}\theta^{10}$ &$d=6$&$ -2\times8$  & closed B-model\cr
$\lambda^7\theta^9$& $d=7$&$ -16$ &Top. M-theory \refs{\GerasimovYX}\cr
\noalign{\smallskip\hrule\vskip .5mm \hrule}\cr
}}$$
\narrower
{\bf Table:} {\ninerm This table lists the single state of highest ghost number for the pure spinor cohomology $H^{(*)}(Q|p.s.)$ for ``open" models and in $H^{(*)}(Q_{L}|p.s.)\otimes H^{(*)}(Q_{R}|p.s.)$ for the ``closed" models.
This state is used for defining the measure of integration for the pure spinor tree-level 
amplitudes. The dimension is the one of the target space once boundary conditions on the fermionic variables are enforced.
The last column lists the name of the theory: in $d=3$ and $d=7$ we have ``open" models and in $d=6$  we have ``closed" models. Finally the ghost anomaly of the model is the sum of the 
ghost number and fermionic number of the highest state for ``open" models, or half of it for the left and right sector for ``closed" models.}
\endgroup
\bigskip

\bigskip
After this analysis have been performed a new interesting paper 
appeared  \DijkgraafTE\ on the archive. There several forms of Form Theories of Gravity in 6 and lower dimensions 
are studied and their lifting to a  7 dimensional topological M theory. Our result is a rather strong piece of evidence 
for a topological M-theory at {\it the quantum level\/}. Indeed, we focused on the relation between the insertions 
needed to saturated the path integral measure for ghosts  
and Grassmann variables that yielded the dimension of the target space 
theory which seems to point out that there is a relation between observables in topological M theory and the  observables of physical M-theory. 

We believe that the present framework gives a complementary
view on twisted topological models, where the difficult part is to find the
original $N=(2,2)$ superstring  model from where they originate. For example, 
in  \ShatashviliZW, a worldsheet analysis is performed, but a derivation of this model 
by considering a topological version of the superstring theory on a $G_{2}$ manifold is still missing. 

\bigskip
The paper is organized as follows. In sec.~2, we present 
a prescription for higher loop contributions to 11d supergravity 
corrections by means of worldline methods. Then we discuss the  
different insertions needed to reabsorb the zero modes and 
we demonstrate that, at two loops, the zero mode saturation selects 
the term $\nabla^{4} R^{4}$ term of the effective action. In sec. 3, we 
show that, by choosing a suitable gauge fixing for the picture changing 
operators (the correspondent gauge parameters are parametrized by 
a spinor $C_{A}$ and the 2-form $B_{MN}$), the prescription given in sec. 2 can 
be reduced to the 10d superparticle prescription given in \multi. In sec. 4, the 
relation between the ghost number of the tree level measure and a 
corresponding topological model is exploited. We conclude in section 5 with a dictionary between the pure spinors supersparticle approach of this
paper with topological string and M theory. The appendix contains some proof of identities of the main text.


\newsec{Higher loop amplitudes  for  pure spinor superparticle formalism}

We briefly recall some ingredients of the multiloop formalism 
for pure spinor superstrings constructed in \multi\ 
and extended to 11d superparticle in \AGV.

By the analysis pursued in there, we recall that  at 
tree and one-loop  the amplitude prescription has a suitable number of unintegrated vertex operators. However, for  $g \geq 2$
only integrated vertex operators,  denoted by $\int d\tau V(\tau)$ with $\tau$ the world-line coordinate, 
are needed. The  pure spinor formulation is based on the following  conjugated pairs of variables $(\theta^{A}, p_{A})$ 
and $(\lambda^{A}, \omega_{A})$ where $\lambda^{A}$ is constrained by \ePs.

 The fermionic variables $\theta^{A}$ have 32 components realized as 
$\theta^{A}=(\vartheta^{a},\tilde\vartheta^{\tilde a})$ with $a,\tilde a=1,\cdots,16$ for the ten 
dimensional case or as a
32-component Majorana
spinor in eleven dimensions, and 
the pure spinor $\lambda^{A}$, satisfying~\ePs, has 11 complex components in ten dimensions or 
23 complex components in eleven dimensions.
The conjugated variables $p_{A}$ and $w_{A}$ have zero modes at higher-loop $g$ given by $32\times g$ for $p_{A}$
(in the following we will work with the independent 
field $d_{A}= p_{A}+ \cdots$) and $11\times g$ or $23\times g$
for $w_{A}$ in 10 and 11 dimensions respectively.\foot{For the 10d case this is justified by the fact that $\theta^{A}$ is a periodic fermion of conformal weight 0
and $p_{A}$, its conjugated momentum, has conformal weight 1. Applying the Riemann-Roch theorem gives that $\# (\theta)_{0}- \# (p)_{0}=1-g$, where
each component of $\theta^{A}$ has one zero mode $\#(\theta)_{0}=1$, giving that each components of $p_A$ has $g$ zero modes $\#(p)_{0}=g$.}
  For saturating the bosonic 
ghosts, we need in the path integral measure a corresponding number 
of Dirac delta function to soak up their zero modes. This can be 
done by respecting the BRST invariance (and the 
decoupling of BRST exact operators) by introducing the 
picture changing operators \refs{\BerkovitsBF}
\eqn\ePCO{\eqalign{
Z_{B} &= \{Q,\Theta\left(B_{MN} \, (\l \G^{MN}w)\right)\}=
B_{MN} (\l \G^{MN} d)\,  \delta\left(B_{MN} (\l \G^{MN}w)\right)\,, \cr
Z_{J} &=\{Q,\Theta\left(\l w\right)\}=( \l^{A} d_{A})\, \delta(\l^{A} w_{A})\,,\cr
Y_{C} &=C_{A} \theta^{A} \delta(C_{A} \lambda^{A})
}}
where $B_{MN}$ and $C_{A}$ are gauge fixing parameters, and $\Theta(x)$ is the Heavyside step function. 
There will be needed as much insertion of $Z_{B,J}$ as the number of components for $w_{A}$.
The parameter  $B_{MN}$ can be chosen in such a way that no normal ordering 
is needed in the expression for $Z_{B}$. Another ingredient needed 
is the picture changed anti-ghost $b_{B}$, which satisfies\foot{In the superparticle limit the $b$-field is 
a density. In string theory this quantity is the density formed by the inner product between the Beltrami differential such that
$\int \, b(t)=\int \mu_{\bar z}{}^{z}(t)\, b_{zz}$.}
\eqn\eDefB{
\{Q, b_{B}\} = Z_{B} T
}
 where $T$ is the stress energy tensor 
(for the superparticle $T = P^{M}P_{M}$ and $b_{B} = \Theta(B_{MN} 
\l \G^{MN} w) T$ see \AGV\ or \OdaBG\ for more comments). 
The number of insertion of  $b_{B}$-anti-ghost in the multiloop amplitude is the number of its zero modes given by 
\eqn\eCG{
c(g) = 7 (g-1)\ .
}
The path integral measure for $\l^{A}$ and for the conjugate 
$w_{A}$ are symbolically given by $[{\cal D}\lambda]_{+16}$ and 
$[{\cal D}N]_{-16}$ where the superscript indicates the ghost charge 
(their complete expressions are given in \AGV). We denote 
by $\tau_{i}$ the Schwinger parameters. 

Finally, 
the $g$-loop N-point correlation function is given by 
\eqn\gloops{\eqalign{
{\cal A}^{g}_{N}& = 
\int{\cal D}\lambda \, d^{32}\theta \, \prod_{i=1}^{g} 
{\cal D}N_{i}\, {\rm d}^{32}d_{i} 
\prod_{j=1}^{c(g)} \int \, dt_{j}\,
 b_{B}(t_{j})\, \times \cr
&\times \prod_{k=c(g)+1}^{22\,g} Z_{B_{k}}\, \prod_{l=1}^{g} Z_{J_{l}} 
\prod_{m=1}^{23} Y_{C_{m}}\, \prod_{n=1}^{N} 
\int d\tau_{n} V^{(0)}_{n}(\tau_{n})
}}
As at one-loop (see \AGV) we can count the zero modes.
\item{$\triangleright$} The $\lambda$-ghost number: The measure $[{\cal D}\lambda]$ contributes to $+16$, 
each $[{\cal D}N]$ factors to $-16$, the $Z_{B,J}$ collectively contributes to $23g - c(g)$ and $Y_{C}$ to $-23$, for a total of
\eqn\eLcount{
16 - 16 g + 23g - c(g) - 23 = (23-16) (g-1) - c(g) = 7 (g-1) - c(g)
}
which is zero for \eCG.
\item{$\triangleright$} We have to saturate the 32 zero modes for the $\theta^{A}$. We have $23$ of them from the
$Y_{C}$, so we should get $9$ of them from the vertex operators.   
 \item{$\triangleright$} We have $32g$ zero modes for the $d_{A}$ which have to be soaked up by the $23g-c(g)$
 from the $Z_{B,J}$, we have $2N-M$ zero modes from the vertex operators,  if $M$ counts the number of zero modes 
 for the $N^{MN}=(\lambda \Gamma^{MN}w)$. 
For the $d$'s coming from the insertions $b_{B}$, 
we have to use the engineering dimension discussed in 
\multi\ and we found that $c(g)\ b_{B}$ contribute to 
$8 c(g)/3 + 4 M/3$. For a non vanishing amplitude there should be enough $d$ zero modes coming from the $b_{B}$
insertion giving
\eqn\edzm{
{5\over 3} \, c(g) + {M\over 3}+ 2N \geq 9g
}
 \bigskip
 
Any multiloop prescription should agree with the non-renormalisation theorem in ten
\refs{\GreenTV,\BerkovitsEX,\rfIengo} 
and  eleven dimensions \GreenKV\ that states, for instance, that $R^{4}$ is not renormalized above one-loop and that the
  four gravitons amplitude contributes to at least $\nabla^{4}R^4$ from two-loop and higher.\foot{The results of \GreenKV\ 
point the fact that the $D^4R^4$ is as well not renormalized by higher-loop amplitudes, but this result  is not a consequence of supersymmetry alone. For instance
the absence of corrections from the 3 loops amplitude would be obtained after integration over the moduli and summing all the superparticle diagrams.}
These theorems are consequences of supersymmetry therefore accessible by zero modes counting. We will show
the number of zero modes~\eCG\ for the $b_{B}$-field is the {\sl only} value  compatible with the  $R^4$ non-renormalisation
theorem.

 \bigskip

We consider four gravitons scattering ($N=4$ in \gloops) at $g\geq 2$ loop order. The $R^4$ non-renormalisation theorems stipulate
that from two-loop the four gravitons  amplitude  contributes to the eleven dimensions effective action to at least
 $\int d^{11}x \nabla^{4} R^{4}$ to
 where 
a suitably contraction of the covariant derivatives $\nabla_{M}$ and 
the Riemann tensor $R_{MNPQ}$ is understood. Recalling that the integrated vertex operators for the graviton
have the structure \refs{\multi,\AGV}
$$
V^{(0)} = \int \, dt\, \left(\cdots + {\cal M}^{MN} {\cal M}^{PQ}\,  {\cal R}_{MNPQ}+\cdots \right)\, e^{ik\cdot X}
$$
where we introduced the Lorentz generator
${\cal M}^{MN}= (d\Gamma^{MN}\theta)+ (\lambda \Gamma^{MN}w)$ and the superfield
${\cal R}_{MNPQ}(x,\theta) = R_{MNPQ}(x) + 
\t^{2} \nabla R_{MNPQ}(x) + \cdots$, 
we only need  the following structure from the vertex operators  (see as well \multi)
\eqn\eVert{
\prod_{i=1}^{4} V^{(0)}_{n}(\tau_{n}) \sim 
(d\G^{M_{i}N_{i}}\theta   )\prod_{r=1}^{7} (\l \G^{P_{i}Q_{i}} w) \ {\cal R}_{M_{i}N_{i}P_{i}Q_{i}}(x,\theta)
}
The $\theta^{A}$ coordinate zero modes counting showed that  9 $\theta$'s should comes from the vertex operator
part \eVert\  which implies that 8 $\theta$'s have to be extracted from the curvature superfields and the expression contributes
to four derivatives. With $N=4$ and $M=7$ we can check that \edzm\ is always satisfied for $g\geq 2$.
\newsec{Reduction to 10d}
 
Now, since this seems to give the correct counting for all loop amplitude, 
we would like to provide also an heuristic argument to 
support the number $c(g)=7(g-1)$ as the correct number of insertions 
of $b_{B}$ by comparing \gloops\ with type IIA superstring 
amplitudes. We recall that the superstring amplitudes (for $g>1$)
are computed by the prescription  of \multi
\eqn\gloopsstrings{
{\cal A}^{g}_{N} = 
\int
{\cal D}\lambda_{L} \, d^{32}\theta_{L} \, {\cal D}\lambda_{R} \, d^{32}\theta_{R} \,\prod_{i=1}^{g} 
{\cal D}N_{i,L}\, {\rm d}^{32}d_{i,L}  \prod_{i=1}^{g} 
{\cal D}N_{i,R}\, {\rm d}^{32}d_{i,R} 
}
$$
\Big| \prod_{j=1}^{3(g-1)} \int \, dz_{j}\,
(\mu| b_{B,L})(z_{j})  
\prod_{k=3(g-1)+1}^{11\,g} Z_{B_{k}}\,  
\prod_{m=1}^{22} Y_{C_{m}} \,\Big|^{2} \prod_{n=1}^{N} 
\int d^{2}z_{n} V^{(0)}_{n}(z_{n},\bar z_{n})
$$
where $L/R$ refers to the left- and right-mover sectors of the 
superstrings. In the following we will focus on the superparticle limit of this amplitude.\foot{
 In this limit
the counting of moduli is the same as it can be understood from the plumbing fixture procedure.
Namely adding a loop to a vacuum superparticle loop diagram requires 3 parameters: two for the
position of each insertion point (the punctures) and one for the length of the line connecting the two punctures.
The amplitudes are then constructed by distributing the vertex operators on the internal lines of the
vacuum diagram.
}

 The measure, the picture changing operators and the insertions (except the vertices) 
can be factorized into left and right-parts. 
There the usual counting of moduli $6(g-1)$ (the number of moduli for 
a punctured Riemann surface) leads to $6(g-1)$ insertions of $b_{B}$'s. They 
are folded with the Beltrami differentials and each of $b_{B}$ carries one picture changing operator
 $Z_{B}$. 
Notice also that the number of 
picture lowering operators $Y_{C}$ soak up correctly  the 
22 zero modes for left and right-pure spinors $\l_{L/R}$. 
The following relation $\left(2\, N^{mn} - \delta^{mn}\, J \right)\, (\gamma_{m})_{ab}= 0$ valid at the 
classical level where $N^{mn}=(\lambda \gamma^{mn}w)/2$  and $J=\lambda w$ states that one can trade the ghost current $J$
for one Lorentz generator. We make the same choice in the definition of the multiloop amplitude in 11d \gloops.
 This will make connection between the superstring prescription \gloopsstrings\ and the 11d prescription \gloops\ clearer.

Reducing the superstring to superparticle, it is
easy to show that the above prescription is still valid and provide 
the correct results for radiative corrections to the four gravitons scattering 
at two loops.  
The difference between the 11 dimension superparticle and 
the 10 dimensions N=2 superparticle can be seen directly by counting 
the number of $b_{B}$ insertions, since \gloopsstrings\ has $6(g-1)$ insertion
when the 11d superparticle needs $6(g-1)+1$ insertions.\foot{In the superparticle limit there is no
Riemann-Roch theorem. There is no Riemann-Roch theorem as well for a 3d membrane world-volume theory without any boundaries.
Thinking the superparticle prescription as a limit of  superstring amplitudes, one can contemplate the possibility
of higher spin ghosts. The Riemann-Roch theorem would require a non integer spin $9/4$ ghost system, 
which does not seem realistic.} 

 The supplementary zero
modes arises when relaxing the  constraints $\l \G^{11} \l =0$  which is the
eleven
dimensions implementation of the  condition $b_{0}^{-}=0$ \refs{\BerkovitsUC,\AGV}.
Using the Fierz identity\foot{We use the following notations:  $G\equiv 0$ for identities true independently
of any constraints and  $G=0$ for constraints. For instance 
$$
(\lambda \Gamma_{MN} \Gamma^{P}\lambda) (\lambda \Gamma_{MN}\lambda)
\equiv 2\, (\lambda \Gamma_{M}\lambda)(\lambda \Gamma^{MN}\lambda)= 0
$$
where the equality  a consequence of~\ePs.}  $(\G^{MN})_{(AB} (\G_{N})_{CD)} \equiv 0$,
we can see that 
\eqn\noteC{
(\l\G^{MN} \l)\, (\l \G_{N}  \l) \equiv 0 
}
 from which it follows
\eqn\noteD{\eqalign{
(\l \G^{11 \, n} \l)\, (\l \G_{n}  \l) &\equiv 0\,, \cr
(\l \G^{m n}\l)\, (\l \G_{n}  \l) +  (\l \G^{m \, 11}\l)\,(\l\G_{11}  \l) &\equiv 0\,, 
}}
 where $m,n = 0, \dots, 9$. Imposing the pure spinor 
 constraint $\l \G^{m} \l=0$ for $m=0,\cdots, 9$, the first equation is
automatically solved and the second  implies  either $\l \G^{11} \l =0$ or 
 $\l \G^{11} \G^{m} \l =0$. 
 The pure spinor condition in 11d requires 
 that $\l \G^{11} \l =0$, but if we relax this condition we automatically 
 get the second option $\l \G^{11} \G^{m} \l =0$.
Using the chiral decomposition of the pure spinor 
 $\lambda^{A}=(\lambda_{L}^{\alpha},\lambda_{\tilde\alpha,R})$ these two equations are
 \eqn\noteE{\eqalign{
 \l^{\a}_{L} \g^{m}_{\a\b} \l^{\b}_{L} + 
\l_{\tilde\a,R} \g^{m,\tilde\a\tilde\b} \l_{\tilde\b,R} &= 0\cr
 (\l^{\alpha}_{L}\,\l_{\a,R})\, \l^{\a}_{L} \g^{m}_{\a\b} \l^{\b}_{L}&=0\cr
 }}
The choice $\l^{\a}_{L} \g^{m}_{\a\b} \l^{\b}_{L}=0$ corresponds to pure spinor conditions in 10d for
Type IIA superstring (the left and right pure spinors have opposite chirality) found in \refs{\BerkovitsFE}.

Performing this reduction the 23 components of the 11d pure spinor decompose according 
$\lambda=(\lambda_{L},\lambda_{R},\rho_{\lambda})$ where 
$\lambda_{L,R}$ are the 11 components of 10d pure spinors of \refs{\BerkovitsUC} and $\rho_{\lambda}$ is an extra  scalar component arising
 from the  rescaling $(\lambda_{L},\lambda_{R})\to (\rho_{\lambda}\, \lambda_{L},\rho^{-1}_{\lambda}\,\lambda_{R})$
  preserving the 11d pure spinor \ePs\ constraints. Likewise for the conjugated ghost $w= (w_{L},w_{R},\rho_{w})$.
The measures  constructed in \AGV\ decomposes as 
\eqn\eDLambda{\eqalign{
[{\cal D}^{23}\lambda]_{+16}&= [{\cal D}^{11}\lambda_{L}]_{+8} \wedge 
[{\cal D}^{11}\lambda_{R}]_{+8}\wedge [{\cal D}\rho_{\lambda}]_{+0}\cr
[{\cal D}^{23}w]_{-16}&= [{\cal D}^{11}w_{L}]_{-8} \wedge 
[{\cal D}^{11}w_{R}]_{-8}\wedge [{\cal D}\rho_{w}]_{+0}\cr
}}
The amplitude~\gloops\ has $2\times(11g - 3(g-1))+1$ insertions of the picture raising operators $Z_{B}$ which is one more than for the superstring amplitude
\gloopsstrings, likewise the the number of picture lowering operators $Y_{C}$. 
But the 11d multiloop
amplitude has $6(g-1)+ (g-1)$ insertions of  $b$-field. The extra $g-1$ $b$-fields and the extra 
$Z_{B}$ are exactly the number needed for saturating the $g$ zero modes for $\rho_{w}$.

\bigskip
\noindent $\triangleright$ {\sl The cohomology for the relaxed constraint}

When relaxing the constraint $\l \G^{11} \l =0$, the BRST operator for the 11d superparticle 
 $Q = \l^{A} d_{A}$ 
 is no longer nilpotent  since $Q^{2} = P_{11}  \l \G^{11} \l$.
 For $P_{11}\neq 0$, we can anyway 
 obtain a nilpotent BRST operator by adding a new pair of ghost 
 fields $(c,b)$ with the commutation relation $\{b,c\} = 1$ such that  
 \eqn\mA{
  Q_{M} = Q + c P_{11} - {1\over 2} b \l \G^{11} \l\,.
  } 
  is now nilpotent since $\{Q_{M}, P_{11}\}=0$. 
  An operator/state in the cohomology of $Q$ depends on the space-time coordinates $x^{M}=(x^{m},x^{11})$
  and the pure spinor ghost $\lambda^{A}$, and an operator/state in the cohomology for $Q_{M}$ depends as well on the $c$ ghost.
   In order 
 to prove the equivalence between the cohomology of the original 
 BRST operator and the new one $Q_{M}$, we observe that 
 given a vertex operator $U^{(n)}(x^{M},\lambda)$ of a given ghost number $n$, 
 in the constrained cohomology 
 $\{Q, U^{(n)} \} = \l \G^{11}\l W^{(n-1)}$ where $W^{(n-1)}(x^{M},\lambda)$ is an 
 auxiliary vertex operator with ghost number $n-1$. Acting again 
 with the BRST operator from the left, one gets 
 $\l \G^{11} \l \left(\p_{11} U^{(n)} - \{Q_{M},  W^{(n-1)}\}\right) = 0$. And since 
 $\l \G^{11} \l$ is non-vanishing we conclude that 
 $\{Q_{M}, W^{(n-1)}\} = \p_{11} U^{(n)}$ (notice that we cannot 
 add a second term proportional to $\l \G^{11} \G^{m} \l$ since this quantity vanishes because 
 we assumed that  $\l \G^{11} \l \neq0$ in~\noteD). 
 Thus, we can construct the new vertex operator 
 \eqn\mB{
 U_{M}^{(n)}(x^M,\lambda, c) =  U^{(n)}(x,\l)  - c W^{(n-1)}(x,\l)
 }
 which satisfies $\{Q_{M}, U^{(n)}_{M} \} =0$.

The amplitude is well defined as long as there is enough insertions of $\delta(w_{A})$.  General considerations  \refs{\Belop} on picture changing operators
 ensure that  the generic form of a picture raising operator is $Z_{B}=\{Q,\Theta(B^{A} w_{A})\}$ and of a picture lowering
 operator is $Y_{C}= C_{A} \theta^{A} \, \delta(C_{A}\lambda^{A})$ and that the amplitude is independent
 of the gauge fixing parameters $B^{A}$ and $C_{A}$.
 In order to perform the reduction 
of the 11d superparticle multiloop prescription to the 10d prescription we have to choose appropriately the parameters
$B_{MN}$ and $C_{A}$ in the picture lowering and raising operators.
We choose the  gauge  parameters  $B_{MN}$ with the Lorentz indices along the ten dimensional directions
$B_{mn}$ with $m,n=0,\cdots, 9$  such that
$B_{MN} (\lambda\Gamma^{MN}w)= B_{mn} \, [(\lambda_{L}\gamma^{mn}w_{L})+(\gamma_{R}\gamma^{mn}w_{R})]$.
And we  make a different choice for the gauge fixing constants appearing in the `extra' picture raising and lowering operators
\eqn\eBElf{\eqalign{
Z_{11}&=\{Q, \Theta(w\Gamma^{11}w)\} =(w \Gamma^{11}d)\, \delta(w\Gamma^{11}w)\cr
Y_{11}&= \lambda \G^{11}\theta\, \delta(\lambda \G^{11}\lambda)\cr
\{Q,b_{11}\}&= Z_{11} \, T\Longleftrightarrow  b_{11} = \Theta(w\G^{11}w) \, T\ .\cr
}}
First of all we remark that $Z_{11}$ and $Y_{11}$ still have ghost number $+1$ and $-1$ respectively. These operators are in fact taking care of the 
zero modes for the scalar ghost component $\rho_{\l}$ and $\rho_{w}$ appearing in \eDLambda. The choice of the 
gauge parameter in $Z_{11}$, breaks the gauge symmetry
of $w_L$ and $w_R$ generated by the 10 pure spinor constraints. However,
the variation is cancelled by the delta function of the remaing PCO as explained below.

We have to notice the following properties: 
the combinations $\hat\gamma=\l^{\a}_{L}\l_{\a,R}$ and $\hat b=w_{\a L} w^{\a}_{R}$ have ghost number 
$+2$ and $-2$, they are commuting and scalar combination of the pure spinor 
ghost fields and their conjugates. Moreover, the combinations 
$\t^{\a}_{L} \l_{\a R} + \l^{\a}_{L} \t_{\a R}$ and $w_{\a L} d^{\a}_{R} + d_{\a L} w^{\a}_{R}$ 
have ghost number $+1$ and $-1$, they are anticommuting and they are also 
scalars. Let us denote the first two combinations as $\hat \gamma$ and $\hat \beta$, 
and the second pair as $\hat c$ and $\hat b$. Then we observe that the 
BRST varations of those fields are 
\eqn\BRSTnew{
Q \hat c = \hat\gamma\,, ~~~~  Q \hat\gamma=0\,,~~~~
Q \hat \beta = \hat b\,, ~~~~ Q\hat b = P_{m} \Big(\l_{L} \g^{m} w_{R} + w_{L} \g^{m} \l_{R}\Big)
}
The last transformation implies that $Q$ is not nilpontent on the field $\hat b$. However, 
if the field $\hat b$ is inserted in the correlation functions, there are the picture changing 
operator $Z_{B}$ containing the delta function $\delta(B_{MN} \l \G^{MN} w)$. 
By choosing $B_{m 11} = P_{m}$, the variation of $\hat b$ vanishes 
(changing the gauge parameters $B_{MN}$ is a BRST exact operation and 
the amplitude will not change under it). 
This allows us to view the quartet $\hat c, \hat b$ and $\hat \gamma$ and $\hat \beta$  as a 
topological quartet with an effective BRST charge $\hat Q=\hat b \hat \gamma$. This system 
decouples from the rest of the theory when reducing the amplitude from 11d to 10d. As 
a further confirmation of this, we notice that for such simple topological model, one 
can construct the picture changing operators (know also the picture operator in 
 \refs{\dvvtop,\VerlindeKU,\DistlerAX}) 
 $\hat c \delta(\hat \gamma)$ which is BRST invariant 
 (but not BRST exact) and the $\hat b \delta(\hat \beta) = \{Q, \Theta(\hat\beta)\}$ 
 which is the picture raising operator. Those picture changing operator 
 obtained by 
 the gauge fixing in \eBElf. 
  The insertions of $\hat c \delta(\hat \gamma)$ and $\prod_{k=1}^{g-1} \hat b \delta(\hat \beta)$ 
  in the amplitudes can be 
 established by observing that this system corresponds to Liouville theory with a given background 
 charge \DistlerAX. This is a first step to have a derivation of the higher genus expansion 
 of the amplitudes in \multi\ and in the present paper. 

With these choices the multiloop amplitude  \gloops\ can be rewritten as the 10d prescription with the factorized expression for the 
\eqn\eExtra{\eqalign{
&\int {\cal D}\rho_{\lambda} \,\prod_{i=1}^g {\cal D}(\rho_{w})_{i}
\int dX_{11} 
\prod_{i=1}^{g}d(P_{11})_{i}
\,  Z_{11}\,  Y_{11}\, \prod_{i=1}^{g-1}b^{i}_{11}\times\cr
&\left|\int [{\cal D}^{11}\lambda]_{+8}  [{\cal D}^{11g}w]_{+8} \, \prod_{i=1}^{11g} Z_{B^{i}}\, \prod_{j=1}^{11} Y_{C}\right|^2 \, \int {\cal V}\cdots \int {\cal V}
}}
which is equivalent to the multiloop prescription given in \refs{\multi} with the replacement of $g$ of the $Z_{B^{i}}$ by  $Z_{J^{i}}$. 

\bigskip
\noindent
$\triangleright$ {\sl First case $P_{11}=0$ :  the  perturbative string amplitudes } 

In this case the BRST charge $Q$ is nilpotent and is the sum $Q_{L}+Q_{R}$ of
the BRST charge for the left and right movers for the superstring. All the states in the Hilbert space are independent of $(X_{11},P_{11})$ and the 
ghost $(c,b)$. Therefore the first line in \eExtra\ factorizes completely and we are left with the perturbative superstring multiloop amplitude given in \refs{\multi}.

 \bigskip
 \noindent
 $\triangleright$ {\sl Second case $P_{11}\neq 0$:  Non perturbative contributions}
 
 For constant $P_{11} = M$, $Q$ 
is the BRST charge for a D0-brane \AG\ 
 where $M$ is its mass.  We showed earlier the  equivalence between the cohomology of $Q$ and $Q_{M}$.
In the case of a compactification on a circle along the 11$^{th}$ dimension, one 
has  
$\p_{11} U^{(n)}_{M} = {k\over R}  U^{(n)}_{M}$ where $k$ is an integer and 
$R$ is the radius of the circle $S^{1}$ of the compactification, so the loop amplitude prescription \gloops\ gives perturbative and non-perturbative amplitudes (with D0-branes)
for type IIA. Even for external states independent of $X_{11}$ and the value of $P_{11}$, the intermediate states running the loops will carry a D0-brane charge giving rise to non-perturbative corrections 
as computed in \refs{\GreenAS}. 


\newsec{Relation between ghost number and dimension}

In the present section, we propose some pieces of evidence pointing out some relations between 
the tree level measure for the supersymmetric models (quantized in the pure spinor 
formalism) and corresponding topological theories.

The pure spinors approach in 10 (respectively in  11 dimensions) gives rise to $N=1$ super Yang--Mills (respectively supergravity) equation of motions in 10d 
(respectively in eleven dimensions), but we show that by an appropriate choice of boundary condition on the  fermionic variables $\theta$, open string topological model, as well as 
A/B (closed) string topological model and the 7d topological model of \refs{\GerasimovYX} and be derived.
 
\subsec{10d, the tree level measure and open topological models}

The relation seems to point out that to the N=1 10d open superstring is characterized by a ghost number 3 measure,
 this number has led to the construction of a string field theory-like action 
\BerkovitsRB\ of the form (where we neglect for the moment the interactions and also  all the complications of the BV formalism by restricting our 
attention to ghost number one, for a more general situation 
see for example \ZwiebachIE)
\eqn\openSFT{
S_{SYM} = 
{\rm Tr}\Big\langle {\cal U}^{(1)} Q_{o} \, {\cal U}^{(1)} \Big\rangle + \cdots
}
for 10d $N=1$ super Yang--Mills theory. 
To define the vertex operator and the fields, we started from 
superstring type IIB and we identify on a D9-brane the 
field as $\t_{L} = \t_{R} \equiv \t$, $\l_{L} = \l_{R} \equiv \l$, 
$d_{z\a} = d_{\bar z\a}$ and $w_{\a z} = w_{\a \bar z}$. 
This corresponds to  a specific choice of boundary conditions 
and they implies that $Q_{L} = Q_{R} \equiv Q_{o}$.
For a more generic situation we refer to  \BI.  
The ghost number of the vertex operator  ${\cal U}^{(1)}$ is one and it 
contains the physical fields \refs{\BerkovitsZK}
\eqn\eUone{
{\cal U}^{(1)}= {1\over 2}\, (\lambda\gamma^m\t) \, a_{m}(x,\theta) + {i\over 12} (\t\g^{mnp}\t) (\l\g_{mnp}\chi) + \cdots
}
The  bracket $\langle\cdot,\cdot\rangle$ is 
computed with the measure $\int d\mu_{5}^{(3)} W^{(3)}_{5} =1$ where
\eqn\meA{
W^{(3)}_{5} = 
\l \g^{m_{1}} \t \,\l \g^{m_{2}} \t \,\l  \g^{m_{3}} \t \,
\t \g_{m_{1} m_{2} m_{3}} \t\,.
}
This measure factor is defined uniquely by the fact that in ten dimensions for the pure spinor $\lambda$ satisfying \ePs, 
and the Fierz identity
\eqn\eFI{
(\lambda\gamma_{m})_{\alpha} (\lambda \gamma^{mn_{1}\cdots n_{4}}\lambda)=0
}
implies that  (see Appendix~A)
\eqn\eFII{\eqalign{
\epsilon_{m_{1}\cdots m_{r}}{}^{n_{1}\cdots n_{d-r}}\, (\lambda \gamma^{m_{1}}\theta)\cdots (\lambda\gamma^{m_{r}} \theta)&=0
 \quad {\rm for} \quad r\geq 6\cr
 (\lambda \gamma^{m_{1}}\theta)\cdots (\lambda\gamma^{m_{5}}\theta)&=
 -{1\over 4}\,  (\lambda \gamma^{m_{1}\cdots m_{5}}\lambda)\, W_{5}^{(3)}
}}
The first formula states that only five  $c^m=\l \gamma^{m}\t$ are linearly independent, and the second states that on the constraint \ePs\ 
$W_{5}^{(3)}$ is the volume density. The $c^{m}$ are anti-commuting variables\foot{These correspond to the 
twisted fermions for the topological sigma model on the world-sheet 
 \WittenZZ.} with ghost number $+1$.
 The open string BRST charge $Q_{o}$ reads
 \eqn\eQo{
 Q_{o}= \lambda^{\a} d_{\a} = \l^{\a} p_{\a} -{1\over 2} c^{m} P_{m} +{1\over 4} \, c^m (\t \g_{m}\partial\t)
 }
 \medskip
 \noindent $\triangleright$ {\sl The A-model}

 We now show how to obtain the A-model by projecting the action \openSFT. For this we  restrict ourselves to the  space defined by  
 \eqn\eDeltaTheta{
 \delta_{\theta}^{2} ={1\over 7!} (\t\g_{m_{1}\cdots m_{7}}\t) \, \delta^{(7)}(y) \, dy^{m_{1}}\wedge \cdots\wedge dy^{m_{7}}
 }
 and define the new bracket with the insertion of this $\delta$-function\foot{This constraint amounts to put D-branes boundary conditions on which the superstring ends. The different
 types of D-branes in the pure spinor formalism are studied in \AG\ and they coincide with the one from the usual RNS formulation.}
 \eqn\eNB{
 \langle \cdots  \rangle_{CS} = \int d\mu_{5}^{(3)} \, \delta_{\t}^{2}
 }
 with this definition it is not difficult to see that (see Appendix~A for details)
 \eqn\eTopoOpen{\eqalign{
 {\rm Tr} \left\langle{\cal U}^{(1)} Q_{o} {\cal U}^{(1)} \, \right\rangle_{CS}&= 
  \int d^{10}x\int d\mu_{5}^{(3)}\,\delta_{\t}^{2} \ {\cal U}^{(1)} Q_{o} {\cal U}^{(1)}\cr
  &= \int d^{7}y\, \delta^{(7)}(y) \, \int d^{3}x\,  {\rm Tr}\left(A_{m} \partial_{n}A_{p}+ \chi \gamma_{mnp}\chi\right)\, \epsilon^{mnp}
 }}
 The interaction term can be computed along the same line giving
 \eqn\eTopoInterOpen{
\left\langle {\cal U}^{(1)} {\cal U}^{(1)} {\cal U}^{(1)}\right\rangle_{CS}= \int d^3x \, {\rm Tr}(A \wedge A \wedge A)\ .
 }
Higher point interactions are defined as 
\eqn\eHigher{
 {\rm Tr}\left\langle {\cal U}^{(1)} {\cal U}^{(1)} {\cal U}^{(1)} \, (\int {\cal V}^{(0)})^{n}\right\rangle_{CS}
}
The zero picture vertex operator reads \refs{\BerkovitsZK}
\eqn\eVzero{\eqalign{
{\cal V}^{(0)}&= \partial \theta^{\a} a_{\alpha}(x,\t) + \Pi^{m} a_{m}(x,\t) + d^\a W_{\a} + {1\over 2} N_{mn} F^{mn}\cr
\Pi^{m}&= \partial x^{m} +{1\over2} \t\g^{m}\partial \t\cr
D_{(\a} a_{\b)} &= (\g^m)_{\a\b} \, a_{m};\qquad
D_{\a}W_{\b}= {1\over 4} \, (\g_{mn})_{\a\b}\, F^{mn}
}}
after integrating over the pure spinor $\l$ and the fermions $\t$ one is left with the reduced amplitude
\eqn\eRam{
S_{higher}= \int d^{3}x\, {\rm Tr}\left( A\wedge A \wedge A
 \, \ll  (e^{ik\cdot x})^3 (\int \widehat{\cal V}^{(0)})^{n}\gg\right)}
 where only the part 
$\widehat{\cal V}^{(0)}=\partial x^{m} a_{m}(x) \, e^{ik\cdot x}$
 of the zero ghost picture vertex operator can contribute.
It is important to remark that the gaugino cannot contribute  to this
 interaction term because of the restriction on the number of $\t$s,  and being non dynamical
 it can be integrated out completely.  
 All the higher-point amplitude contains the inverse of the space-time metric $g^{mn}$ and therefore can be scaled away in the limit
 $g_{mn}\to t^2\, g_{mn}$ with $t\to\infty$.

By projecting the 10d pure spinors approach of \refs{\BerkovitsFE} on a 3 dimensional space using~\eNB\ and scaling out the metric, 
we  reproduce  Witten's 3 dimensional Chern-Simons theory \WittenFB\ which is the Chern-Simons theory on $T^{*}M$ is the 
string field theory description of the open topological A model (see for example the review 
 \MarinoUF).
 On the restricted space defined by the constraint \eDeltaTheta, the BRST operator $Q_{o}$ reduces to the de Rham 
differential $d=c^{m}\, \partial_{m}$. 

\bigskip\noindent
$\triangleright$ {\sl The B model}

We also have to take into account the existence of the topological 
B  model. This is characterized by the fact that, unless we restrict 
to a Calabi-Yau manifold, the U(1) charge associated to the ghost number 
is anomalous. We can reproduce the topological B model, by starting 
from closed superstring of type IIB, and we observe the 
the ghost number $(1,0)$ vertex operators of the 
form 
\eqn\topB{
{\cal U}^{(1,0)}_{L} = \l^{\a}_{L} A_{\a}(x,\t_{L},\t_{R})\,, 
}
are BRST closed under $Q_{L}$ if \refs{\BerkovitsZK}
$(\g^{m_{1}\cdots m_{5}})^{\a\b}D_{L,(\a} A_{\b)} =0$
where the superfields $A_{\a}$  depends on 
both of the coordinates 
$\t_{L}$ and $\t_{R}$. This implies that only for $\t_{R}=0$, 
the equations of motion describes the SYM theory on shell. 
However, the combination 
$\left.{\cal U}^{(1,0)}_{L} Q_{L} {\cal U}^{(1,0)}_{L}\right|_{\t_{R}=0}$ inserted into the tree-level path 
integral measure  vanishes, because of the integration over the $\t_{R}$ variables. Therefore, the only way to get a non-trivial result 
we have to insert  $W^{(3)}_{5,R}$ the unique element of $H^{(3)}(Q_{R}|p.s.)$ and the action is
\eqn\topBB{\eqalign{
S_{h} &={\rm Tr}\Big \langle W^{(3)}_{5,R}  
\Big({\cal U}^{(1,0)}_{L} Q_{L} {\cal U}^{(1,0)}_{L} \Big)\Big\rangle_{CS} + 
{\rm Tr}\left\langle W^{(3)}_{5,R}  
\Big({\cal U}^{(1,0)}_{L}  \Big)^3\right\rangle_{CS}+\cdots \,.
}}
We have as well inserted a $\delta_{\t_{L}}^{2}$ in the measure for the left fermions as indicated by the subscript $CS$ on the bracket. 
Notice that the presence of $W^{(3)}_{5,R}$ has two purposes: 
{\it i)} it saturates the ghost charge of the vacuum and {\it ii)}  by inserting 
the vertex $W^{(3)}_{5,R}$, the Grassmann variables $\t_{R}$ are totally 
soaked up, and 
projects ${\cal U}^{(1,0)}_{L} Q_{L} {\cal U}^{(1,0)}_{L}$ on  the space 
$\t_{R}=0$. As for the A model Chern-Simons action, all higher-point amplitudes are
suppressed by scaling away the metric. 

There is a  close analogy with the holomorphic Chern-Simons theory for the topological B model   (here ${\cal M}_{6}$ is 
a Calabi-Yau 3-fold)
\eqn\topBC{
S_{hCS} = \int_{{\cal M}_{6}} \Omega \wedge {\rm Tr}\left(A \bar \partial A + {2\over3}A^3\right) \ .
}
The globally defined holomorphic 3-form $\Omega$ is 
replaced by the scalar (gauge singlet) measure $W^{(3)}_{5,R}$ in \topBB. The latter is needed to compensate 
the ghost current anomaly, in the same way  the presence 
of $\Omega$ is needed in order to compensate the ghost anomaly 
of the topological model \refs{\BookTop}. The   vertex operator $W^{(3)}_{5,R}=c^{m}_{R}c_{R}^n c^{p}_{R}\, (\t_{R}\g_{mnp}\t_{R})$,
with $c^{m}_{R}=\l_{R} \g^{m}\t_{R}$ can be view as defining the holomorphic 3-form with the identification\foot{In the case of RNS, the vertex 
operator $W^{(3)}_{5,R}$ is replaced by $c\p c \p^{2}c e^{2 \phi}$ 
whose interpretation from the target physics is rather obscure. On 
the other side, in pure spinor formulation the explicit super-Poincar\'e 
invariance and the usage of superspace renders the interpretation 
rather transparent.}  
 the anti-commuting ghost $c^m$ with one forms and $\t_{R} \g_{mnp} \t_{R}$ with the 3-form  $\Omega_{mnp}$ of SU(3)-structure manifold. We recall that in the superparticle limit, the variable $\theta$ reduces to its zero mode.

 \subsec{10d, $N=2$, the tree level measure and the  A and B topological models}

For the closed topological A/B models, the situation is very similar. 
Starting from pure spinor superstrings, the tree level measure is 
obtained by duplicating the $W^{(3)}_{5}$ for the left- and right-movers 
that we denote by $W^{(3)}_{5,L}W^{(3)}_{5,R}$. This measure is BRST 
closed and not BRST exact, so it belongs to the cohomology 
$H^{(3)}(Q_{L}| p.s.) \otimes H^{(3)}(Q_{R}| p.s.)$. To construct a 
string field theory model, the vertex operator mush have ghost number 2, 
${\cal U}^{(1,1)}$ and therefore one has to insert an operator $c^{-}_{0}$ to 
construct a kinetic term (see for example \ZwiebachIE)
\eqn\closedK{
 S_{\rm closed} = 
 \langle {\cal U}^{(1,1)} c^{-}_{0} (Q_{L} + Q_{R}) {\cal U}^{(1,1)} \rangle+ \langle{\cal U}^{(1,1)} {\cal U}^{(1,1)} {\cal U}^{(1,1)}   \rangle\,.} 
  However, 
for pure spinor formulation in 10d there is no $c^{-}_{0}$ to construct the kinetic 
term, we will show in the next subsection how this arise by reducing the 11d construction of \refs{\multi,\AGV}.\foot{ In  
 \GrassiIH, the first author proposed an action with an infinite number of 
auxiliary fields (as suggested in 
 \BerkovitsWJ\ and 
\BerkovitsTN) and this points out that it can be replaced by a 
non-local action.} By comparing with the topological model, we have to consider 
closed A/B models whose string field theory description is 
provided in 
\BershadskyCX\ and in 
 \IqbalDS\ (a string field theory for topological A model is 
also covered in 
 \BershadskySR) and 
the action is written in terms of a (1,1) form $A'$
\eqn\KS{
S_{KS} = {1\over 2} \int A' {1\over \partial} \bar \partial A'+ {1\over 6}\int (A'\wedge A')\wedge A'
}   
where the inverse differential operator (well-defined on the massive states 
of the theory) coincides with the ghost field $c^{-}_{0}$, and being $b^{-}_{0} \equiv 
\partial$ and $L_{0} - \bar L_{0} = \Delta - \bar \Delta$ with $\Delta = \partial^{\dagger} \partial$, 
one get that $\{c^{-}_{0}, b^{-}_{0} \} = 1$ on the massive states. Therefore, 
$S_{KS} = (A', c^{-}_{0} Q A')$ where the differential $\bar\partial$ is identified
with the BRST operator.  The same mapping is applied for the A-model of  \refs{\WittenFB,\BershadskySR} with the action 
\eqn\eKG{
S_{KG} = {1\over 2}\int K {1\over d^{c\dagger}} dK +{1\over 6}\int K\wedge K \wedge K
}
where now $b_{0}^{-}\equiv d^{c\dagger}$ and $L_{0}-\bar L_{0}= \Delta - \bar \Delta$ with $\Delta=d^{c\dagger}d$.

The problem to construct a string field theory action for 
closed topological model is very similar to the construction above of string field theory 
for type IIA/B for the full-fledged superstring with pure spinors since there is no $c^{-}_{0}$. 

Notice again the relation between the dimension of the spacetime for the topological 
model and the ghost number of for the level measure and the counting of $b_{B}$ insertion. 
As suggested in \BerkovitsUC, the counting of degrees-of-freedom for N=2 type IIA/B superparticle models 
reveals that there are 8 bosonic degrees of freedom versus 20 fermions degrees-of-freedom. 
Four of the latter are interpreted as coming from $c_{0,L}, c_{0,R}$ and $b_{0,L}, b_{0,R}$ and 
therefore the level matching condition is not automatically implemented. On the 
other side, for 11d superparticle, this naive counting of degrees-of-freedom shows that 
there are 9 bosonic degrees-of-freedom, but only 18 fermion degrees-of-freedom. From the latter 2 of them are 
read as the $b^{+}_{0}$ and $c^{+}_{0}$, 
while $b^{-}_{0}$ and $c^{-}_{0}$ are automatically taken into account.
This seems to suggest that in 11d a string field theory action can be indeed found. 
 
 In the next section, we explain the origin of the $c_{0}^-$ from 11d and as well how the action \KS\ and~\eKG\ can be derived 
 along the line of the previous sections.


\subsec{11d, the tree level measure and Gerasimov-Shatashvili topological model}

We briefly recall some ingredients of 11d pure spinor formalisms. We 
describe the tree level measure (while the all loop amplitude are 
described in the previous section) and we argue that from the string field 
theory action (for the massless fields, so a quantum field theory), 
which was established in \BerkovitsUC\ and extended beyond the kinetic 
term in \AGV\ one can obtain a string field theory action for type IIA string theory. 
The relation with topological  models is seen in the following way: 
from the tree level measure and from higher loop expansion we found that 
the dimension of the spacetime for the corresponding topological model should be 
7. Recently in \GerasimovYX, it was pointed out that there is a description of the closed 
topological model type B (whose string field theory is identified with Kodaira-Spencer 
theory) in term of a local action in one higher dimension. We show that the 
form of the 7d Hamiltonian of \refs{\GerasimovYX} can be indeed guessed from the 
string field theory for the present 11d superparticle description. 

First, we discuss the tree level measure for 11d, then we write the 
supergravity action in a Chern-Simons form, the relation with the
functional by Gerasimov-Shatashvili, and finally we show that reducing from 11d 
to 10d we found that precisely the eleventh component of the pure spinor constraint 
leads to $c_{0}^{-}$ discussed above. 

 $\l^{A}$ denotes a Majorana commuting spinor in 11d, $A=1, \dots, 32$, and 
 it satisfies the 11d pure spinor condition
\eqn\ePS{
\l^{A} \G^{M}_{AB} \l^{B} = 0\,,
}
 with $M=0, \dots 10$ (notice that the ${\rm dim}_{\bf C} \, Spin(10,1) = 32$, and 
 the Majorana condition reduce it to  ${\rm dim}_{\bf R} \, Spin(10,1) = 32$. 
 To solve the pure spinor constraints in 11d with signature $(10,1)$ we
 have to use Dirac complex spinors   $\l^{A}$).  
 $\G^{M}_{AB}$ are $32 \times 32$ symmetric Dirac matrices. 
  Since the $\l^{A} \l^{B}$ is a symmetric bi-spinor it can be 
 decomposed into a basis of Dirac matrices as follows 
 \eqn\noteA{
32\,\l^{A} \l^{B} = 
 \G_{M}^{AB} (\l \G^{M} \l) + 
{1 \over 2!}\, \G_{[MN]}^{AB} (\l \G^{[MN]} \l) + 
{1 \over 5!}\, \G_{[MNPQR]}^{AB} (\l\G^{[MNPQR]} \l)\,, 
}
The first term vanishes thanks to the pure spinor constraint 
and the pure spinor satisfies the Fierz identity
\eqn\eNC{
(\lambda\Gamma^{M})_{A}\, (\lambda \Gamma_{MN}\lambda)=0
}
This Fierz identity implies that zero momentum cohomology of the BRST operator $Q$ with the pure spinor condition stop at $\lambda$-ghost number 7
\eqn\CSA{
W_{9}^{(7)} =
\l \G^{M_{1}} \t  \dots \l \G^{M_{7}} \t  \t \G_{M_{1} \dots N_{7}} \t\,,
} 
and that the eleven dimensions supergravity fields and antifields belong to $H^{(3)}(Q|{p.s.})\oplus H^{(4)}(Q|p.s.)$.

With the measure $\int d\mu_{9}^{(7)}\, W^{(7)}_{9} =1$, one can construct the 
target space action $S_{11d}$ by observing that the vertex operator 
${\cal U}^{(3)}$ contains the supergravity fields and the BRST charge has ghost number 
1. As shown in \BerkovitsUC\ and extended at non-linear level in \AGV\ 
we have\foot{The ellipsis stand for the quartic terms, {\it e.g.} the four fermions terms, 
computed in  \AGVtwo, and for higher-point interactions (4-point and higher).} 
\eqn\sugraeleven{
S_{11d} = 
\langle {\cal U}^{(3)} Q {\cal U}^{(3)} \rangle + \langle {\cal U}^{(3)} [{\cal U}^{(1)}, {\cal U}^{(3)}] \rangle + \cdots\,.
}
As before we restrict the integration by specifying  boundary  conditions with the insertion of 
\eqn\eDeltaEleven{
\delta^{2}_{\t}={1\over 4!}\, (\t\g_{m_{1}\cdots m_{4}}\t) \,\delta^{4}(y)\,dy^{m_{1}}\wedge\cdots\wedge dy^{m_{4}}
}
As before we consider the action
\eqn\eTopoM{
S_{H} = 
\langle {\cal U}^{(3)} Q {\cal U}^{(3)} \rangle_{CS} + \langle {\cal U}^{(3)} [{\cal U}^{(1)}, {\cal U}^{(3)}] \rangle_{CS} + \cdots\,.
}
The vertex operators ${\cal U}^{(3)}$ contains the graviton and the 3-form at order 
$\lambda^3\t^3$ and the gravitino at order $\l^3\t^4$, with the expression  \refs{\BerkovitsUC}
\eqn\exapU{\eqalign{
{\cal U}^{(3)} &=(\lambda\Gamma^{(M}\theta)(\lambda\Gamma^{N)K}\theta)(\lambda\Gamma_{K}\theta)\,
g_{MN}\cr
&+(\l \G^{M} \t)\, (\l \G^{N} \t)\, (\l \G^{P} \t)\, C_{MNP} \cr
&+ (\lambda\Gamma^{M}\theta)\left[(\lambda\Gamma^{N}\theta)(\lambda\Gamma^{P}\theta)
(\theta\Gamma_{NP}\Psi_{M})-(\lambda\Gamma^{NP}\theta)(\lambda\Gamma_{N}\theta)
(\theta\Gamma_{P}\Psi_{M})\right]\,.
 }}
For the interaction term we just need to know that in ${\cal U}^{(1)}$ all the physical fields appear at least at order $\t^2$ \refs{\AGV}. This
forbids any contributions from the interactions and we are left with the {\sl exact} seven dimensional
action for the 3-form
\eqn\gs{
S_{H} = \int d^{7}x \, \left(C \wedge d C +  \Psi_{m_{1}}
 \Gamma^{m_{1}m_{2}}\Psi_{m_{2}} \right)
}
where $C$ is the three form and $d$ is the de Rham differential. As before upon the restriction
imposed by \eDeltaEleven, the BRST operator $Q$ reduced to $c^M\, \partial_{M}$.
Gerasimov and Shatashvili showed that by Hamiltonian reduction how to 
obtain from \gs\ the Kodaira-Spencer theory of \refs{\BershadskyCX} by analyzing the a suitable wave function 
for the path integral. Again the fermion being non-dynamical they can be integrated out from \gs.

\bigskip\noindent
$\triangleright$ {\sl The level matching condition}
 
We now show that from the 11d analysis, we can recover the insertion 
a candidate for $c^{-}_{0}$ confirming the conjecture in \BerkovitsUC. 

The element $W^{(7)}_{9}$ in \CSA\ has ghost number 7 and is of order $\t^{9}$. 
If we relax the constraint $\l \G^{11} \l =0$, then 
\eqn\CS{
Q_{M} W^{(7)}_{9} = (\l \G^{11} \l) W^{(6)}_{8}\,,
}
where $W^{(6)}_{8}$ is the vertex operator to be identified 
with the closed string zero momentum cohomology at highest ghost number. 
By a simple counting, one sees that the BRST differential $Q_{M}$ reduces the 
number of $\t$'s by an unity and therefore $W^{(6)}_{8} \sim \l^{6} \t^{8}$ which does not match the states
 $(\l^6\t^{10})$ in $H^{(3)}(Q_{L}|p.s.)\otimes H^{(3)}(Q_{R}|p.s.)$.
 But, we have also to recall that by eliminating 
the constraint $\l\G^{11} \l = 0$, the number of possible invariants with ghost number seven increases. 
There is another  term of the form $W^{(7)}_{11} \sim \l^{7} \t^{11}$   such that
\eqn\CSC{ 
Q_{M}W^{(7)}_{11}= (\l\G^{11}\l) \, W^{(6)}_{10}\ .
}
Solving this equation at zero momemtum, with $Q_{M}=Q_{L}+Q_{R}$ gives (see Appendix~A for an alternative derivation)
\eqn\CSCsol{
  W^{(7)}_{11} = \left( \l \G^{11} \t \right) W^{(3)}_{5,L} W^{(3)}_{5,R}\,.
}
where  $Q_{L/R} W^{(3)}_{5,L/R} =0$.
This gives the relation between the  tree level measure for 11d and that 
of the type IIA N=2 superparticle
\eqn\Relation{
\langle {\cal U}^{(3)} Q {\cal U}^{(3)}\rangle_{\l \G^{11}\l\neq 0}= \langle  {\cal U}^{(1,1)} c_{o}^{-} (Q_{L}+Q_{R}) {\cal U}^{(1,1)}\rangle
}
with
\eqn\Cnot{
c_{0}^{-} = \l \G^{11}\t\ .
} 
Notice that the factor $\l \G^{11} \t$ 
impose the addition constraint $c^{-}_{0}$ for the level 
matching.

\medskip

 We are finally able to confirm explicitly the conjectured relation 
between the 11d measure needed to write the type IIA string action in a 
covariant way. The K\"ahler two-form of the action~\eKG\ for 
the A model arises from the ghost number 2 element of the cohomology
\eqn\kk{
K={1\over 2}\,\l_{L} \g^{m} \t_{L} \l_{R} \g^{n} \t_{R} (\t_{L} \g_{m} \g_{n} \t_{R})
}
which correspond to a vertex operator of IIA superstring ${\cal U}^{(1,1)}= \l^{\a} \l^{\tilde\a} \, A_{\a\tilde\a}$ with 
constant RR field $P^{\a}_{\hat\b} = \delta^{\a}_{\hat\b} f$ where 
$f$ is a constant coefficient. (This is dual to $\star F_{10}$. 
The potential $F_{10} = d C_{9}$ couples to the $D_{8}$ branes.)
Notice that it is peculiar that this non-propagating degrees-of-freedom of 
the superstring provides here the K\"ahler form.  Inserted in \closedK\ we reproduce the action \eKG.

\medskip

And what about type IIB? As is well known, the 
problem of self-dual 5-form affects the construction of a kinetic term  for 
string field theory in the usual way. However, there are several alternatives: 
one is to use an infinite number of field or non-polynomial 
expressions as we discussed above.  


\newsec{A dictionary}

In the present section, we propose a dictionary between 
pure spinor formulation of superstrings, superparticle and 
supermembranes and topological theories on manifold 
with special holonomies. 

Let us start from the case of open superstring. We found that 
the monomial $W^{(3)}_{5}$, dual to the path integral measure  
on the zero modes, yields the 3-form $\t \g_{mnp} \t$. This 
form resemble the usual calibration for compactification 
of string theory on a space with special holonomy. The 
spinor bilinear $\t \g_{mnp} \t$, built from the $\theta^A$ zero modes, can be 
identified with the 3-form for a 3-fold (a Lagragian submanifold) 
coinciding with its volume form. This provides a dictionary between the 
open superstring with topological A model. If the we consider the 
supersymmetric sector of the heterotic string as the pure 
spinor string theory, we can identify the 3-form 
$\t_{L} \g_{mnp} \t_{L}$ as the homolorphic 3-form. If the 
right moving sector is provided by a topological string on a 
3-fold CY, we can construct a topological B model. 

Let us now consider closed superstring model. In that 
case the volume form is provided by the product 
$W^{(3)}_{5,L} W^{(3)}_{5,R}$. In this case  we 
identified the holomorphic sector of the CY with the 
left moving sector of superstring and vice-versa the anti-holomorphic 
sector with the right movers. In this way we can 
identify the 3-form $\t_{L} \g_{mnp} \t_{L}$ with 
the holomorphic 3-form and $\t _{R}\g_{mnp} \t_{R}$ with 
the antiholomorphic component. Notice that they identify the 
CY space with holonomy $SU(3)$.

In the case of 11d, we have found that the tree level measure 
for superparticle (and for supermembrane) is $W^{(7)}_{9}$ giving the four form $\t \G_{MNPQ} \t$.
 The dimension 
of the manifold is identified by the total ghost number of $W^{(7)}_{9}$ 
(and from the number of $b$-field insertion in the higher-loop formula).
 The four form  $\t \G_{MNPQ} \t$, 
restricted to 7 dimensions is dual to the 3-form which is together 
with the four form provide the complete characterization of the 
$G_{2}$-holonomy space. 

\medskip
The construction of this paper exhibits special states in the pure spinor cohomology associated with
invariant forms characterizing  manifolds of special holonomy $SU(3)$ and $G_{2}$. The vertex operators
 for these forms are part of the
 measure of 
integration of the effective Chern-Simons models, and are crucial for consistency of the model (with the boundary discussed in
the main text). The forms are made from the zero modes of the fermionic coordinates due to
the superparticle approximation, but  a similar construction from a pure spinor formulation of the superstrings \refs{\BerkovitsZK}
 and the supermembrane \refs{\BerkovitsUC} would give non-constant invariant forms (the superparticle is the zero mode approximation
 of the superstring or the supermembrane).
Finally, in order to verify the correctness of the present dictonary, it 
would be interesting to provides also a mapping between the 
amplitudes and try to see which sectors of the correlation functions 
can be indeed computed by the using topogical models.

\newsec{Acknowledgments}

We would like to thank N. Berkovits,  L. Castellani, L. Alvarez-Gaum\'e, L. Magnea, M. Mari\~no, N. Nekrasov,  
R. Russo,  S. Shatashvili  for useful discussions. We thank  S. Theisen for a careful reading of the paper, and Lilia Anguelova for interesting comments.
The work of P.A.G.  is partially funded by NSF grant PHY-0354776. 
P.V. acknowledges the RTN MRTN-CT-2004-503369 and  MRTN-CT-2004-005104 for partial financial support. 

\appendix{A}{Technicalities and proofs of various identities}

\item{$\triangleright$} {\bf The  gaugino part of \eTopoOpen:} For this it is the convienient to use the following representation of the integration measure 
\meA\
\eqn\meAII{
\int d\mu_{5}^{(3)}\, (\lambda \g^{m_{1}}\t) \cdots (\l\g^{m_{3}}\t) (\t \g^{m_{4}\cdots m_{10}}\t) = {1\over 7!}\,\epsilon^{m_{1}\cdots m_{10}}
}
and the following Fierz identity
\eqn\eFI{
(\t \g^{mnp}\t) (\l \g_{mnp}\chi) (\t \g^{rst}\t) (\l\g_{rst}\chi)
\propto   (\t \g^{p_{1}\cdots p_{3}}\t) (\t \g^{p_{4}st}\t) (\l\g_{p_{1}\cdots p_{5}}\l )
 (\chi\g^{stp_{5}}\chi)
}
\break
\item{$\triangleright$} {\bf Proof of the identity \eFII.}

In 10d the pure spinors $\lambda$ satisfies the Fierz identities
\eqn\eFFI{\eqalign{
\lambda_{\alpha}\lambda_{\beta} &= {1\over 16\cdot 5!}\, (\lambda \gamma_{p_{1}\cdots p_{5}}\lambda)\, (\gamma^{p_{1}\cdots p_{5}})_{\alpha\beta}\cr
(\lambda\gamma_{m})_{\alpha}\, (\lambda \gamma^{mn_{1}\cdots n_{4}}\lambda)&=0
}}
Considering 
\eqn\eLT{\eqalign{
16\cdot 5!\, (\lambda \gamma^{m}\theta) (\lambda\gamma^{n}\theta)&= (\lambda\gamma_{p_{1}\cdots p_{5}}\lambda) (\theta \gamma^m \gamma^{p_{1}\cdots p_{5}}\gamma^n \theta)\cr
&= (\lambda\gamma_{p_{1}\cdots p_{5}}\lambda) \left[(\theta \gamma^{mp_{1}\cdots p_{5}n}\theta)
-20 \delta^{mn}_{[p_{1}p_{2}}(\theta \gamma_{p_{3}\cdots p_{5}]} \theta)\right]\cr
}}
we can show that 
\eqn\eBB{\eqalign{
\epsilon_{m_{1}\cdots m_{r}}{}^{n_{1}\cdots n_{10-r}} (\lambda\gamma^{m_{1}}\theta)\cdots (\lambda\gamma^{m_{r}}\theta)&=
-{1\over 16\cdot 5!}(\lambda\gamma_{p_{1}\cdots p_{5}}\lambda) \,  (\lambda\gamma^{m_{1}}\theta)\cdots (\lambda\gamma^{m_{r-2}}\theta)\cr
 \times\left[{8!\over 3!}\,\delta^{p_{1}\cdots p_{5}q_{1}\cdots q_{3}}_{m_{1}\cdots m_{r-2}n_{1}\cdots n_{10-r}} (\theta \gamma_{q_{1}\cdots q_{3}} \theta)\right.
&+\left.20\,\epsilon_{m_{1}\cdots m_{r-2}}{}^{n_{1}\cdots n_{10-r}p_{1}p_{2}}\, (\theta \gamma^{p_{3}\cdots p_{5}} \theta) \right]
}}
The first term vanishes for $r-2> 3$ because of the second identity in \eFFI. The second term vanished because
\eqn\eBBII{\eqalign{
(\lambda\gamma_{p_{1}\cdots p_{5}}\lambda)  \, \epsilon_{m_{1}\cdots m_{r-2}}{}^{n_{1}\cdots n_{10-r}p_{1}p_{2}}&=
 2!\,(\lambda\gamma^{m_{1}\cdots m_{r-2}n_{1}\cdots n_{10-r}}\gamma_{p_{3}\cdots p_{5}}\lambda)\cr
 &= {2!8!\over 5!} (\lambda \gamma^{[m_{1}\cdots m_{r-2}n_{1}\cdots }\lambda)\, \delta^{n_{8-r}\cdots n_{10-r}]}_{p_{3}\cdots p_{5}}
}}
which vanish for $r-2>3$ when plugged back into the second term of \eBB.

\item{$\triangleright$} {\bf Derivation of \CSCsol}

We start from the ten-dimensional left and right measures
\eqn\eTenW{\eqalign{
(\l_{L}\g^{m_{1}}\t_{L})\cdots (\l_{L}\g^{m_{5}}\t_{L}) &= (\l_{L}\g^{m_{1}\cdots m_{5}}\l_{L})\, W^{(3)}_{5,L}\cr
(\l_{R}\g^{m_{1}}\t_{R})\cdots (\l_{R}\g^{m_{5}}\t_{R}) &= (\l_{R}\g^{m_{1}\cdots m_{5}}\l_{R})\, W^{(3)}_{5,R}\ .\cr
}}
Multiplying these two equations using that for $m=0,\dots,9$ $\l \G^{m}\t= \l_{L}\g^{m}\t_{L}+ \l_{R}\g^{m}\t_{R}$ and
 that $\wedge^{r}\, (\l_{L/R}\g^{m}\t_{L/R})= 0$ for $r>5$ and $(\l_{L/R}\g^{[m_{1}\cdots m_{5}}\l_{L/R})(\l_{L/R}\g^{m_{6}\cdots m_{10}]}\l_{L/R})=0$  for  each of the ten-dimensional 
chiral pure spinors, we get that
\eqn\eTenWW{
(\l\G^{m_{1}}\t)\cdots (\l\G^{m_{10}}\t) = (\l\G^{11}\G^{m_{1}\cdots m_{5}}\l) (\l \G^{11} \G^{m_{6}\cdots m_{10}}\l)\, W^{(3)}_{5,L}W^{(3)}_{5,R}
}
Multilplying this equation by  $\l\G^{11}\t$ and Fierzing the $\l$s on the right-hand-side we have 
\eqn\eRE{
\epsilon_{M_{1}\cdots M_{11}}\, (\l\G^{M_{1}}\t)\cdots (\l\G^{M_{11}}\t)\propto  (\l\G^{11}\l)^2 \, (\l\G^{11}\t) \, W^{(3)}_{5,L}W^{(3)}_{5,R}\, .
}
Which gives the vertex operator $W^{(7)}_{11}$ of \CSCsol.

\listrefs

\end